\documentstyle[12pt]{article}

\scrollmode
\begin{document}

\begin{center}
{\Large \bf On a Generalized Kepler-Coulomb System:}

\vspace{0.5cm}
{\Large \bf Interbasis Expansions}
\end{center}

\vspace{1cm}

\begin{center}

{\bf M.~KIBLER}

Institut de Physique Nucl\'eaire de Lyon,

IN2P3-CNRS et Universit\'e Claude Bernard,

43 Boulevard du 11 November 1918,

F-69622 Villeurbanne Cedex, France

\vspace{0.5cm}
{\bf L.G.~MARDOYAN}

Yerevan State University,

Mravyan st.~1,

375049 Yerevan, Armenia

\vspace{0.5cm}
{\bf G.S.~POGOSYAN}

Bogolubov Laboratory of Theoretical Physics,

Joint Institute for Nuclear Research,

141980 Dubna, Moscow region, Russia

\vspace{2cm}

{\bf Abstract}

\end{center}

\vspace{0.3cm}

This paper deals with a dynamical system that generalizes
the Kepler-Coulomb system and the Hartmann system.
It is shown that the Schr\"odinger equation for this
generalized Kepler-Coulomb system can be separated in
prolate spheroidal coordinates. The coefficients of the
interbasis
expansions between three bases (spherical, parabolic and
spheroidal) are studied in detail.
It is found that the coefficients
for the expansion of the parabolic basis in terms of the
spherical basis, and vice-versa, can be expressed through the
Clebsch-Gordan coefficients for the group SU(2)
analytically continued to real values of their
arguments. The coefficients for the expansions
of the spheroidal basis in terms of the spherical
and parabolic bases are proved
to satisfy three-term recursion relations.

\vspace{1 cm}


{\bf Preprint Lycen 9415}. Submitted for publication in International
Journal of Quantum Chemistry.

\newpage
\baselineskip 0.8 true cm

\begin{center}
\section{.~Introduction}
\end{center}

The purpose of the present paper is to further study the
nonrelativistic quantum mechanical system corresponding
to the three-dimensional axially symmetric potential
\begin{equation}
V = \alpha \frac{1} {\sqrt{x^2 + y^2 + z^2} } +
    \beta  \frac{z} {\sqrt{x^2 + y^2 + z^2} } \frac{1}{ x^2 + y^2 } +
    \gamma                                    \frac{1}{ x^2 + y^2 }
\end{equation}
(in Cartesian coordinates),
where $\alpha$, $\beta$ and $\gamma$ are constants such that
$\alpha < 0$ and $\gamma \geq |\beta|$. If $\beta = \gamma = 0$,
we have the ordinary (spherically symmetric) Kepler-Coulomb potential.
In the case where
$\beta    = 0$ and
$\gamma \ne 0$, the potential (1)
reduces to the Hartmann potential
that has been
used for describing axially symmetric
systems like ring-shaped molecules
[1-3] and
investigated from different points of view
in the last decade
[4-20]. The generalized Kepler-Coulomb system corresponding to
$\beta  \ne 0$ and
$\gamma \ne 0$ has been worked out in Refs.~[9,~14,~17-20]. In
particular, the (quantum mechanical) discrete spectrum for the
generalized Kepler-Coulomb system is well known [9,~14,~17],
even for the so-called ($q,p$)-analogue of this system [17].
Furthermore, a path integral treatment of the potential (1)
has been given in
Refs.~[14,~18]. Recently, the dynamical symmetry of the
generalized Kepler-Coulomb system has been studied in
Refs.~[17,~19,~20], the classical motion of a particle moving
in the potential (1) has been considered in Ref.~[19], and the
coefficients connecting the parabolic and spherical bases have
been identified in Ref.~[20] as Clebsch-Gordan coefficients of
the pseudo-unitary group SU$(1,1)$.

The potential $V$ [see Eq.~(1)]
belongs to a set of three-dimensional potentials
systematically investigated in the 1960s
(in connection with the question of accidental degeneracy)
by the late Professor Smorodinsky and
some of his students and/or collaborators
[21-23] and revived, in recent years,
by Evans
[24-26].

The generalized Kepler-Coulomb system turns out to
be interesting from both a classical-mechanical and a
quantum-mechanical point
of view. Indeed, not all the (bounded)
classical trajectories are periodic and the quantum energy
spectrum exhibits accidental degeneracies with respect to the
geometrical group O($2$). (The latter two points are probably
connected.) The accidental degeneracy arises from the fact that
the Schr\"odinger equation for the potential $V$ is separable in
more than one system of coordinates [21-23]. In addition, the
potential $V$ generalizes the Kepler-Coulomb potential, that is
of central importance in quantum chemistry, and, like the
Hartmann potential, may have applications [possibly in a
($q,p$)-deformed form or in a supersymmetric form] in chemical
physics for systems presenting a line of singularity.

A further motivation for the present work is the following.
It has been
recognized for a long time that the Schr\"odinger equation and
the Hamilton-Jacobi equation for the potential (1) can be
separated in spherical polar and
             parabolic rotational
coordinates [23]. However, to the best of our knowledge, it has
never been questioned, even in the recent papers of Refs.~[9,~14,~17-20],
if these equations are also
separable in an additional coordinate system as the system
of prolated   ellipsoidal and
            hyperboloidal coordinates. This is
actually the case and the originality of our work is
mainly concerned with this separability in prolate spheroidal
coordinates.

The system of spheroidal coordinates constitute a natural
system for
investigating many problems in mathematical physics
(see Ref.~\cite{b20} and references cited therein). In quantum
mechanics, the spheroidal coordinates
play an important role because they are appropriate in describing
the behavior of a charged particle in the field of two Coulomb centres.
The distance $R$ between the centres is a dimensional
parameter characterizing the spheroidal coordinates. These
coordinates are changed into
spherical and parabolic coordinates
as $R\rightarrow 0$ and $R\rightarrow\infty$, respectively, if the
positions of one Coulomb centre and of the charged particle are
fixed
when taking the limits. In this sense, the spheroidal
coordinates are more general than the spherical and parabolic ones.
This explains the interest of spheroidal coordinates in
various domains (astrophysics, plasma physics, theory of the
chemical bond, etc.). Such an interest is well documented in quantum
chemistry, for instance in the study of two-center systems like
H$_2^+$ and
H$_2  $ (see Refs.~[28-34] for a nonhexhaustive list
of papers showing the interest of spheroidal coordinates
in quantum chemistry).

The paper is organized as follows.
In Sections 2 and 4, we describe the
spherical and parabolic bases, for
the generalized Kepler-Coulomb system, in a way adapted
to the introduction of interbasis expansions.
In Section 3, we prove an additional orthogonality
property for the spherical radial wavefunctions
of given orbital quantum number $l$.
In Section 5, by using the property of bi-orthogonality
of the spherical basis, we calculate the coefficients
of interbasis expansions between spherical and parabolic
bases.
In Section 6, we construct the spheroidal basis of
the potential (1) in terms of the superposition of the spherical
and parabolic bases.
Finally, we prove in the appendix that the
bi-orthogonality property (discussed in
Section 3) of the spherical radial
wavefunctions is a consequence of the
accidental degeneracy of the discrete
spectrum for the generalized Kepler-Coulomb
system.

\begin{center}
\section{.~Spherical Basis}
\end{center}

By introducing the nonnegative constants $c_{i}$
$(i=1,2)$ such that
\begin{equation}
\beta  = c_2 - c_1,\qquad
\gamma = c_1 + c_2,
\end{equation}
the potential $V$ [see Eq.~(1)] can be
rewritten in the P\"{o}schl-Teller form
\begin{equation}
V = \alpha \frac{1}{r} +
    c_1    \frac{1}{2r^2 \cos^2 \frac{\theta}{2}} +
    c_2    \frac{1}{2r^2 \sin^2 \frac{\theta}{2}},
\end{equation}
in spherical coordinates ($x = r \sin \theta \cos \varphi$,
                          $y = r \sin \theta \sin \varphi$,
                          $z = r \cos \theta$). The Schr\"odinger equation
\begin{equation}
H\Psi = E\Psi
\end{equation}
for the potential (3) may be solved by
searching for a wavefunction in the form
\begin{equation}
\Psi(r,\theta,\varphi) = R(r) Z(\theta,\varphi).
\end{equation}
This leads to the two coupled differential equations
\begin{equation}
\frac{1}{\sin \theta} \frac{\partial}{\partial \theta}
\left( \sin \theta \frac{\partial Z}{ \partial \theta}  \right) +
\frac{1}{\sin^2 \theta} \frac{\partial^2 Z}{\partial \varphi^2} -
\left( \frac{c_1}{\cos^2 \frac{\theta}{2}} +
\frac{c_2}{\sin^2 \frac{\theta}{2}} \right) Z = - A Z,
\end{equation}
\begin{equation}
\frac{1}{r^2} \frac{d}{dr} \left( r^2 \frac{dR}{dr} \right)  -
\frac{A}{r^2} R + 2 \left(E - \frac{\alpha}{r} \right) R =
0,
\end{equation}
where $A$ is a separation constant in spherical coordinates.
[We use a system of units for which the reduced mass $\mu$
and the Planck constant $h$ satisfy $\mu = h / (2 \pi) = 1$.]

The solution of Eq.~(6) is easily found to be
\begin{equation}
Z(\theta,\varphi) \equiv
Z_{lm}(\theta,\varphi;\delta_1,\delta_2)
  =  N_{lm}(\delta_1,\delta_2)
(1 + \cos \theta)^{\frac{|m|+\delta_1}{2}}
(1 - \cos \theta)^{\frac{|m|+\delta_2}{2}}
P_{l-|m|}^{(|m|+\delta_2,
            |m|+\delta_1)} (\cos \theta) {\rm e}^{im\varphi},
\end{equation}
where $l \in {\bf N}$,
      $m \in {\bf Z}$,
$\delta_i = \sqrt{m^2 + 4c_i} - |m|$ (for $i=1,2$)
and $P_n^{(\alpha, \beta)}$ denotes a Jacobi
polynomial.
Furthermore, the separation constant $A$  is quantized as
\begin{equation}
A = \left(l + \frac{\delta_1 + \delta_2}{2} \right)
\left(l + \frac{\delta_1 + \delta_2}{2} + 1 \right).
\end{equation}
The normalization constant $N_{lm}(\delta_1,\delta_2)$
in (8) is given  (up to a phase factor) by
\begin{equation}
N_{lm}(\delta_1,\delta_2) = \frac{(-1)^{\frac{m-|m|}{2}}}{2^{|m|}}
\sqrt{\frac{(2l+\delta_1+\delta_2+1)(l-|m|)!
\Gamma(l+|m|+\delta_1+\delta_2+1)} {2^{\delta_1+\delta_2+2} \> {\pi}
\>
\Gamma(l+\delta_1+1) \>
\Gamma(l+\delta_2+1)}}.
\end{equation}
The angular wavefunctions $Z_{lm}$ [see Eq.~(8)] shall be referred to as
ring-shaped functions. These ring-shaped functions
generalize the functions studied by
Hartmann [1-3]
in the case $\beta = 0$
(i.e., $\delta_1 = \delta_2$).
Due to the connecting formula
\cite{b22}
\begin{equation}
\left( \lambda + \frac{1}{2} \right)_n C_n^{\lambda}(x) = (2\lambda)_n
P_n^{\left( \lambda - \frac{1}{2}, \lambda - \frac{1}{2} \right)}
(x)
\end{equation}
between the Jacobi polynomial $P_n^{(\alpha,\beta)}$ and the
        Gegenbauer polynomial $C_n^{\lambda       }$, the case
$\delta_1 = \delta_2 = \delta$ yields
$$
Z_{lm}(\theta,\varphi;\delta,\delta) =
(-1)^{\frac{m-|m|}{2}} 2^{|m|+\delta}
\Gamma \left( |m|+\delta+\frac{1}{2} \right)
\sqrt{\frac{(2l+2\delta+1)(l-|m|)!}{4\pi^2 \Gamma(l+|m|+2\delta+1)}}
$$
\begin{equation}
\cdot \> (\sin \theta)^{ |m|+\delta               }
             C_{l-|m|}^{ |m|+\delta + \frac{1}{2} }
                     (\cos \theta) {\rm e}^{im\varphi},
\end{equation}
a result already obtained in Ref.~\cite{b23}.
[In Eq.~(11), $(x)_n$ stands for a Pochhammer symbol.]
The case $\delta = 0$ (i.e.,
         $\beta  =
          \gamma = 0$) can be treated
by using the connecting formula
\begin{equation}
P_l^{|m|}(x) = \frac{(-2)^{|m|}}{\sqrt{\pi}} \Gamma
\left( |m|+\frac{1}{2} \right) (1 - x^2)^{\frac{|m|}{2}}
C_{l-|m|}^{|m|+\frac{1}{2}}(x)
\end{equation}
between the Gegenbauer polynomial    $C_n^{\lambda}$
and the associated Legendre function $P_l^{|m|    }$
\cite{b22}. In fact for $\delta = 0$, Eq.~(12) can be reduced
to
\begin{equation}
Z_{lm}(\theta,\varphi; 0,0) = (-1)^{\frac{m+|m|}{2}}
\sqrt{       \frac {2l+1    }
                   {4\pi    }
\>           \frac {(l-|m|)!}
                   {(l+|m|)!}
}
P_l^{|m|} (\cos \theta)
{\rm e}^{im\varphi},
\end{equation}
an expression that coincides with the usual (surface) spherical
harmonic $Y_{lm}(\theta,\varphi)$ (e.g., see \cite{b24}). The
surface spherical harmonics thus arise as particular cases of more
general functions, viz., the ring-shaped functions.

Let us go now to the radial equation. The
introduction of (9) into (7) leads to
\begin{equation}
\frac{1}{r^2} \frac{d}{dr} \left(r^2 \frac{dR}{dr} \right) -
\frac{1}{r^2}
  \left( l + \frac{\delta_1+\delta_2}{2}     \right)
  \left( l + \frac{\delta_1+\delta_2}{2} + 1 \right) R +
2 \left( E - \frac{\alpha}{r}                \right) R = 0,
\end{equation}
which is reminiscent of the radial equation for the hydrogen
atom except that the orbital quantum number $l$ is replaced
here by
$l + (1/2)(\delta_1 + \delta_2)$. The solution of (15) for the
discrete spectrum is
\begin{equation}
R(r) \equiv R_{nl}(r;\delta_1,\delta_2) = C_{nl}(\delta_1,\delta_2)
(\varepsilon r)^{l + \frac{\delta_1+\delta_2}{2}}
{\rm e}^{-\frac{\varepsilon}{2} r}
{_1F}_1(-n+l+1, 2l+\delta_1+\delta_2+2; \varepsilon r),
\end{equation}
where $n \in {\bf N} - \{ 0 \}$. In Eq.~(16), the
normalization factor $C_{nl}(\delta_1, \delta_2)$
reads
\begin{equation}
C_{nl}(\delta_1,\delta_2) = \frac{2(-\alpha)^{ \frac{3}{2} }}
{\left(n + \frac{\delta_1+\delta_2}{2}\right)^2}
\frac{1}{\Gamma(2l+\delta_1+\delta_2+2)}
\sqrt{\frac{\Gamma(n+l+\delta_1+\delta_2+1)}{(n-l-1)!}}
\end{equation}
and the parameter $\varepsilon$ is defined by
\begin{equation}
\varepsilon = - \frac{2\alpha} { n + \frac{\delta_1+\delta_2}{2} }.
\end{equation}
The eigenvalues $E$
are then given by
\begin{equation}
E \equiv E_n = -\frac{\alpha^2}{2 \left( n + \frac{\delta_1+\delta_2}{2}
\right)^2}, \qquad n = 1, 2, 3, \cdots,
\end{equation}
in agreement with Ref.~\cite{b16} (see also Refs.~[9,~17]).
Equations (16) to (19)
can be specialized
to the cases $\delta_1 = \delta_2 = \delta$ (with
$\delta\neq 0$ or $\delta = 0$). In the limiting case
$\delta = 0$, we recover the familiar results for
hydrogenlike atoms.

\begin{center}
\section{.~Bi-Orthogonality of the Radial Wavefunctions}
\end{center}

The radial wavefunctions $R_{nl}$ [see Eq.~(16)]
satisfy the orthogonality relation
\begin{equation}
I_{nn'} = \int_{0}^{\infty} R_{n'l}(r;\delta_1,\delta_2)
R_{nl}(r;\delta_1,\delta_2)r^{2} dr = \delta_{nn'},
\end{equation}
which is necessary for the normalization of the total
wavefunction
$\Psi_{nlm} (r, \theta, \varphi ; \delta_1, \delta_2)$
given by
Eqs.~(5), (8), and (16).
In addition to the condition (20), we have also the
following orthogonality condition
\begin{equation}
J_{ll'} = \int_{0}^{\infty}
R_{nl'}(r;\delta_1,\delta_2)
R_{nl }(r;\delta_1,\delta_2) dr
= \frac{2\alpha^2}{\left(n + \frac{\delta_1+\delta_2}{2} \right)^3}
\, \frac{1}{2l + \delta_{1} + \delta_{2} + 1}
\, \delta_{ll'}.
\end{equation}
This new relation shall prove useful when dealing
with the interbasis expansions in Section 5. The proof of Eq.~(21)
is as follows.

In the integral appearing in (21), we replace $R_{nl}$
and $R_{nl'}$ by their expressions (16).
Then, we take the hypergeometric function
${_1F}_1 (-n+l+1, 2l+\delta_1 + \delta_2 + 2 ; \varepsilon r)$ in
(16) as the finite sum
\begin{equation}
{_1F}_1 \left(-n+l+1, 2l+\delta_1 + \delta_2 + 2 ; \varepsilon r \right) =
\sum_{s=0}^{n-l-1}
\frac{(-n+l+1)_s}
{(2l + \delta_1 + \delta_2 + 2)_s}
\frac{(\varepsilon r)^s}{s!}
\end{equation}
and we perform the integration term by term
with the help of the formula \cite{b25}
\begin{equation}
\int_{0}^{\infty} {\rm e}^{-\lambda x} x^\nu {_1F}_1(\alpha, \gamma; kx) \, dx
=
\frac{\Gamma(\nu+1)}{\lambda^{\nu+1}} \,
{_2F}_1 \left( \alpha, \nu+1, \gamma ; \frac{k}{\lambda} \right).
\end{equation}
By using
\begin{equation}
{_2F}_1(a, b, c ; 1) =
\frac{\Gamma(c)\Gamma(c-a-b)}{\Gamma(c-a)\Gamma(c-b)},
\end{equation}
we arrive at
\begin{eqnarray}
J_{ll'} & = & \frac{2\alpha^2}
{\left( n + \frac{\delta_1+\delta_2}{2} \right)^3} \,
\frac{\Gamma(l+l'+\delta_1+\delta_2+1)}{\Gamma(2l+\delta_1+\delta_2+2)}
\, \sqrt{\frac{1}{(n-l-1)! (n-l'-1)!} \>
         \frac{\Gamma(n + l + \delta_1 + \delta_2 + 1)}
              {\Gamma(n + l'+ \delta_1 + \delta_2 + 1)}}  \nonumber\\
 &  & \cdot \> \sum_{s=0}^{n-l-1}
\frac{(-n+l+1)_s}{s!} \,
\frac{(l+l'+\delta_1+\delta_2+1)_s}{(2l+\delta_1+\delta_2+2)_s} \,
\frac{\Gamma(n-l-s)}{\Gamma(l'-l-s+1)}.
\end{eqnarray}
By introducing the formula \cite{b22}
\begin{eqnarray}
\frac{\Gamma(z)}{\Gamma(z-n)} = (-1)^n \frac{\Gamma(-z+n+1)}{\Gamma(-z+1)}
\end{eqnarray}
into (25), the sum over $s$
can be expressed in terms
of the
${_2F}_1$
Gauss hypergeometric
function of argument 1.
We thus obtain
$$
J_{ll'} =
\frac{2\alpha^2}{\left(n + \frac{\delta_1+\delta_2}{2}\right)^2}
\frac{1}{l+l'+\delta_1+\delta_2+1}
$$
\begin{equation}
\cdot \> \sqrt{\frac {(n-l -1)!}
                     {(n-l'-1)!}
               \frac {\Gamma(n+l +\delta_1+\delta_2+1)}
                     {\Gamma(n+l'+\delta_1+\delta_2+1)} }
\frac{1}{\Gamma(l-l'+1)\Gamma(l'-l+1)}.
\end{equation}
Equation (21) then easily follows from (27)
since $[\Gamma(l-l'+1)\Gamma(l'-l+1)]^{-1}
= \delta_{ll'}$.

The result provided by Eq.~(21) generalizes the one for the
hydrogen atom \cite{b21}. Indeed, orthogonality properties
similar to (21) hold for the hydrogen atom system and the
harmonic oscillator system in $f$-dimensional spaces ($f \ge 2$)
\cite{b21}. Such unusual orthogonality properties are connected
to the accidental degeneracies of the energy spectra for these
systems. In this connection, the
property (21) is a consequence of the
accidental degeneracy (with respect to the orbital
quantum number $l$) of the discrete energy spectrum for the
generalized Kepler-Coulomb system under consideration~; this
point is further studied in the appendix.

\begin{center}
\section{.~Parabolic Basis}
\end{center}

In the parabolic coordinates
\begin{eqnarray}
x = \sqrt{\mu\nu} \cos \varphi, \quad
y = \sqrt{\mu\nu} \sin \varphi, \quad
z = \frac{1}{2} (\mu - \nu),    \quad
0 \leq \mu, \nu < \infty,       \quad
0 \leq \varphi  < 2\pi,
\end{eqnarray}
the potential $V$ reads
\begin{eqnarray}
V = \alpha \frac{2}{    \mu+\nu } +
    c_1    \frac{2}{\mu(\mu+\nu)} +
    c_2    \frac{2}{\nu(\mu+\nu)}.
\end{eqnarray}
By looking for a solution of the Schr\"odinger
equation (4) for this potential in the form
\begin{equation}
\Psi(\mu,\nu,\varphi) = f_1(\mu) f_2(\nu) {\rm e}^{im\varphi},
\end{equation}
we obtain the two coupled equations
\begin{equation}
\frac{d}{d\mu} \left( \mu \frac{d f_1}{d\mu} \right) +
\left[ \frac{E}{2} \mu - \frac{(|m|+\delta_1)^2}{4 \mu} + B_1 \right]
f_1  =  0,
\end{equation}
\begin{equation}
\frac{d}{d\nu} \left( \nu \frac{d f_2}{d\nu} \right) +
\left[ \frac{E}{2} \nu - \frac{(|m|+\delta_2)^2}{4 \nu} + B_2 \right]
f_2  =  0,
\end{equation}
where the two separation constants $B_{1}$ and $B_{2}$
obey $B_1 + B_2 = - \alpha$. (In the whole paper, we use $\Psi$
to denote the total wavefunction whatsoever the coordinate
system is. The wavefuctions $\Psi$ in spherical, parabolic and
prolate spheroidal coordinates are then distinguished by the
corresponding quantum numbers.)
By solving (31) and (32), we get the normalized wavefunction
\begin{eqnarray}
\Psi(\mu,\nu,\varphi) \equiv
\Psi_{n_1 n_2 m}(\mu,\nu,\varphi;\delta_1,\delta_2) =
\frac{{\varepsilon}^2}
{\sqrt{-8\alpha}}
f_{n_1 m} (\mu; \delta_1)
f_{n_2 m} (\nu; \delta_2)
\frac{{\rm e}^{im\varphi}}{\sqrt{2\pi}},
\end{eqnarray}
where
$$
f_i(t_i)\equiv
f_{n_{i}m}(t_i;\delta_i)  = \nonumber\\
\frac{1}{\Gamma(|m|+\delta_i+1)}
\sqrt{\frac{\Gamma(n_i +|m|+\delta_i+1)}{n_i!}}
$$
\begin{eqnarray}
\cdot \> \left( \frac{\varepsilon}{2} t_i \right) ^{\frac{|m|+\delta_i}{2}}
{\rm e}^{-\frac{\varepsilon}{4} t_i}
\> {_1F}_1 \left( -n_i,|m|+\delta_i+1; \frac{\varepsilon}{2} t_i \right),
\end{eqnarray}
with $i=1,2$ ($t_1\equiv\mu$ and $t_2\equiv\nu$). In Eq.~(34), we have
\begin{eqnarray}
n_i  =  -\frac{|m|+\delta_i+1}{2} + \frac{2}{\varepsilon}
B_i, \qquad (i = 1,2).
\end{eqnarray}
Here again $m\in {\bf Z}$ in order to ensure the univaluedness
of $\Psi_{n_1 n_2 m}$. In addition, $n_{i} \in {\bf N}$ (for
$i=1,2$) in order that $\Psi_{n_1 n_2 m}$ be in $L^{2}( {\bf R}^3 )$.
Therefore, the quantized values of the energy $E$
are given by
(19) where now the quantum number $n$ is $n = n_1 + n_2 + |m| + 1$,
a number that parallels the principal quantum number
of the hydrogen atom in parabolic coordinates.

The results (33) to (35) agree with the corresponding
ones of Ref.~\cite{b19} derived by making use of the
Kustaanheimo-Stiefel transformation (see also Ref.~\cite{b18}).

Following Kibler and Campigotto \cite{b19}, we can obtain a further
integral of motion besides $E$. In parabolic coordinates,
this integral corresponds to the hermitean operator
\begin{equation}
X = \frac{2}{\mu+\nu} \left[ \mu\frac{\partial}{\partial\nu}
\left( \nu\frac{\partial}{\partial\nu} \right) -
\nu\frac{\partial}{\partial\mu} \left( \mu\frac{\partial}{\partial\mu}
\right) \right] + \frac{\mu-\nu}{2\mu\nu}
\frac{\partial^2}{\partial\varphi^2} +
\frac{2c_1\nu}{\mu(\mu+\nu)} - \frac{2c_2\mu}{\nu(\mu+\nu)} -
\alpha \frac {\mu-\nu} {\mu+\nu}.
\end{equation}
The eigenvalues of $X$ are
\begin{equation}
B_1 - B_2 = - \alpha\frac{n_1 - n_2 + \frac{\delta_1-\delta_2}{2}
}{n + \frac{\delta_1+\delta_2}{2}}.
\end{equation}
In Cartesian coordinates, the operator $X$ can be rewriten as
\begin{equation}
X  =  z\left( \frac{\partial^2}{\partial x^2} +
\frac{\partial^2}{\partial y^2} \right) -
x \frac{\partial^2}{\partial x \partial z} -
y \frac{\partial^2}{\partial y \partial z} - \frac{\partial}{\partial z}
- \alpha \frac {z} {r} + c_1\frac{r-z}{r(r+z)} -
c_2\frac{r+z}{r(r-z)},
\end{equation}
so that it immediately follows that $X$ is connected to the
$z$-component $A_z$ of the Laplace-Runge-Lenz-Pauli vector via
\begin{equation}
X = A_z + c_1\frac{r-z}{r(r+z)} - c_2\frac{r+z}{r(r-z)}
\end{equation}
and coincides with $A_z$ when $\beta = \gamma = 0$.

\begin{center}
\section{.~Interbasis Expansion between Parabolic and Spherical
Bases}
\end{center}

The connection between the spherical ($r, \theta, \varphi$)
                       and parabolic ($\mu,  \nu, \varphi$)
coordinates is
\begin{equation}
\mu = r(1 - \cos\theta),\qquad
\nu = r(1 + \cos\theta),\qquad
\varphi ( {\rm parabolic} ) =
\varphi ( {\rm spherical} ).
\end{equation}
For a given subspace corresponding to a fixed
value of $E_n$, we can write the parabolic wavefunction
(30) in terms of the spherical wavefunctions (5) as
\begin{equation}
\Psi_{n_1 n_2 m}(\mu,\nu,\varphi;\delta_1,\delta_2) =
\sum_{l=|m|}^{n-1} W_{n_1 n_2 m}^l(\delta_1,\delta_2)
\Psi_{nlm}(r,\theta,\varphi;\delta_1,\delta_2).
\end{equation}
By virtue of Eq.~(40), the left-hand side of (41) can be rewritten
in spherical coordinates. Then, by substituting $\theta=0$
in the so-obtained equation and by taking into account
that
\begin{equation}
P_n^{(\alpha,\beta)}(1) = \frac{(\alpha+1)_n}{n!},
\end{equation}
we get an equation which depends only on the variable $r$.
Thus, we can use the orthonormality relation (21)
on the orbital quantum numbers $l$.
This yields
\begin{equation}
W_{n_1 n_2 m}^l(\delta_1,\delta_2) =
\frac{(-1)^ {\frac {m-|m|} {2} } }
{\Gamma(|m|+\delta_1+1) \Gamma(2l+\delta_1+\delta_2+2)} \>
E_{n_1 n_2}^{lm} \>
K_{n   n_1}^{lm},
\end{equation}
where
$$
E_{n_1 n_2}^{lm} =
\sqrt{(2l+\delta_1+\delta_2+1)(l-|m|)!\Gamma(l+\delta_1+1)}
$$
\begin{equation}
\cdot \> \left[ \frac{\Gamma(n_1+|m|+\delta_1+1)
\Gamma(n_2+|m|+\delta_2+1)
\Gamma(n+l+\delta_1+\delta_2+1)}{(n_1)!(n_2)!(n-l-1)!\Gamma(l+\delta_2+1)
\Gamma(l+|m|+\delta_1+\delta_2+1)} \right] ^{1/2}
\end{equation}
and
\begin{equation}
K_{nn_1}^{lm} = \int_{0}^{\infty} x^{l+|m|+\delta_1+\delta_2} {\rm e}^{-x}
\> {_1F}_1(-n_1  , |m|+\delta_1+1         ; x)
\> {_1F}_1(-n+l+1, 2l+\delta_1+\delta_2 + 2 ; x) \, dx.
\end{equation}
To calculate the integral $K_{nn_1}^{lm}$, it is sufficient to write the
confluent hypergeometric function
${_1F}_1(-n_1, |m|+\delta_1+1 ; x)$
as a series, to integrate according to (23)
and to use the formula (24) for the summation of the Gauss
hypergeometric function ${_2F}_1 (a, b, c ; 1)$. We thus obtain
$$
K_{nn_1}^{lm} = \frac{(n-|m|-1)!\Gamma(2l+\delta_1+\delta_2+2)
\Gamma(l+|m|+\delta_1+\delta_2+1)}{(l-|m|)!\Gamma(n+l+\delta_1+\delta_2+1)}
$$
\begin{eqnarray}
\cdot \> {_3F}_2 \left\{
\begin{array}{l}
-n_1,  -l+|m|, l+|m|+\delta_1+\delta_2+1 \\
|m|+\delta_1+1,   -n+|m|+1
\end{array}
\biggr| 1 \right\}.
\end{eqnarray}
The introduction of (46) into (43) gives
$$
W_{n_1 n_2 m}^l(\delta_1,\delta_2) = (-1)^{ \frac{m-|m|}{2} } \>
\frac{(n-|m|-1)! \Gamma(l+|m|+\delta_1+\delta_2+1)}
{(l-|m|)!\Gamma(|m|+\delta_1+1)\Gamma(n+l+\delta_1+\delta_2+1)}
$$
\begin{equation}
\cdot \> {_3F}_2 \left\{
\begin{array}{l}
-n_1,  -l+|m|,   l+|m|+\delta_1+\delta_2+1 \\
|m|+\delta_1+1,  -n+|m|+1  \\
\end{array}
\biggr| 1 \right\} E_{n_1 n_2}^{lm}
\end{equation}
and owing to (44) we end up with
$$
W_{n_1 n_2 m}^l(\delta_1,\delta_2) =
\sqrt{\frac{(2l+\delta_1+\delta_2+1)
\Gamma(n_1+|m|+\delta_1+1)\Gamma(n_2+|m|+\delta_2+1)
}{(n_1)!(n_2)!(n-l-1)!
(l-|m|)!}}
$$
$$
\cdot \> (-1)^{\frac{m-|m|}{2}}
\frac{(n-|m|-1)!}{\Gamma(|m|+\delta_1 + 1)}
\sqrt{\frac{\Gamma(l+\delta_1+1)
\Gamma(l+|m|+\delta_1+\delta_2+1)}
{\Gamma(l+\delta_2+1)
\Gamma(n+l+\delta_1+\delta_2+1)}}
$$
\begin{equation}
\cdot \> {_3F}_2 \left\{
\begin{array}{l}
-n_1,  -l+|m|,   l+|m|+\delta_1+\delta_2+1 \\
|m|+\delta_1+1,  -n+|m|+1  \\
\end{array}
\biggr| 1 \right\}
\end{equation}
that constitutes
a closed form expression for the interbasis coefficients.

The next step is to show that the interbasis coefficients
(48) are indeed a continuation on the real line of the
Clebsch-Gordan coefficients for the group SU(2).
It is known that the Clebsch-Gordan coefficient
$
C_{a \alpha ;
   b \beta}^{c\gamma} \equiv
\langle{a\alpha b\beta | a b c \gamma}\rangle
$
can be written as \cite{b24}
$$
C_{a \alpha ;
   b \beta}^{c\gamma} = (-1)^{a-\alpha}
\delta_{\gamma,\alpha+\beta}(a+b-\gamma)!(b+c-\alpha)!
$$
$$
\cdot \> \left[ \frac{(2c+1)(a+\alpha)!(c+\gamma)!}
{(a-\alpha)!(b-\beta)!(b+\beta)!(c-\gamma)!(a+b+c+1)!(a+b-c)!(a-b+c)!
(b-a+c)!} \right] ^{1/2}
$$
\begin{equation}
\cdot \> {_3F}_2 \left\{
\begin{array}{l}
-a-b-c-1, -a+\alpha, -c+\gamma \\
-a-b+\gamma,  -b-c+\alpha  \\
\end{array}
\biggr| 1 \right\}.
\end{equation}
In view of the formula \cite{b26}
\begin{equation}
{_3F}_2 \left\{
\begin{array}{l}
s, s', -N \\
t', 1-N-t  \\
\end{array}
\biggr| 1 \right\} =
\frac{(t+s)_N}{(t)_N}
\> {_3F}_2 \left\{
\begin{array}{l}
s, t' - s', -N \\
t', t+s  \\
\end{array}
\biggr| 1 \right\},
\end{equation}
equation (49) can be rewritten in the form
$$
C_{a \alpha ;
   b \beta}^{c\gamma} = (-1)^{a-\alpha}
\delta_{\gamma,\alpha+\beta}
\left[ \frac{(2c+1)(b-a+c)!(a+\alpha)!(b+\beta)!(c+\gamma)!}
{(a-\alpha)!(b-\beta)!(c-\gamma)!(a+b-c)!(a-b+c)!(a+b+c+1)!} \right] ^{1/2}
$$
\begin{equation}
\cdot \> \frac{(a+b-\gamma)!}{(b-a+\gamma)!}
\> {_3F}_2 \left\{
\begin{array}{l}
-a+\alpha, c+\gamma+1, -c+\gamma \\
\gamma-a-b,  b-a+\gamma+1  \\
\end{array}
\biggr| 1 \right\}.
\end{equation}
By comparing Eqs. (51) and (48),
we finally obtain
\begin{equation}
W_{n_1 n_2 m}^l(\delta_1,\delta_2) = (-1)^{n_1 + \frac{m-|m|}{2}}
C_{\frac{n+\delta_2-1}{2}, \frac{|m|+n_2-n_1+\delta_2}{2};
\frac{n+\delta_1-1}{2}, \frac{|m|+n_1-n_2+\delta_1}{2}}^{l+
\frac{\delta_1+\delta_2}{2}, |m| + \frac{\delta_1+\delta_2}{2}}.
\end{equation}
Equation (52) proves that the coefficients for the expansion of
the parabolic basis in terms of the spherical basis are nothing
but the analytic continuation, for real values
of their arguments, of the SU(2) Clebsch-Gordan coefficients.

 The inverse of Eq. (41), namely
\begin{equation}
\Psi_{nlm}(r,\theta,\varphi;\delta_1,\delta_2) =
\sum_{n_1=0}^{n-|m|-1} \widetilde W _{nlm}^{n_1}(\delta_1,\delta_2)
\Psi_{n_1 n_2 m}(\mu,\nu,\varphi;\delta_1,\delta_2),
\end{equation}
is an immediate consequence of the orthonormality property of
the SU(2) Clebsch-Gordan coefficients. The expansion
coefficients in (53)
are thus given by
\begin{equation}
\widetilde W_{nlm}^{n_1}(\delta_1,\delta_2) = (-1)^{n_1 + \frac{m-|m|}{2}}
C_{\frac{n+\delta_2-1}{2}, \frac{n+\delta_2-1}{2} - n_1;
\frac{n+\delta_1-1}{2}, n_1+|m|-\frac{n-\delta_1-1}{2}}^{l+
\frac{\delta_1+\delta_2}{2}, |m| + \frac{\delta_1+\delta_2}{2}}
\end{equation}
and  may  be  expressed  in  terms of the
${_3F}_2$ function through (49) or (51).

To close this section, it should be mentioned that (52) and
(54) generalize the well-known result, corresponding to
$\delta_1 = \delta_2 = 0$, for the interbasis expansion
between parabolic and spherical bases obtained in
Refs.~[41-44] in the case of the hydrogen atom. Furthermore,
by taking $\delta_1 = \delta_2 \ne 0$ in (52) and (54),
we recover our former result \cite{b7}
for the Hartmann system.

\begin{center}
\section{.~Prolate Spheroidal Basis}
\end{center}

\vspace{0.3cm}
\noindent {\it 6.1. Separation in Prolate Spheroidal Coordinates}

We now pass to the prolate spheroidal coordinates
\begin{equation}
x = \frac{R}{2} \sqrt{(\xi^2-1)(1-\eta^2)} \cos \varphi,\quad
y = \frac{R}{2} \sqrt{(\xi^2-1)(1-\eta^2)} \sin \varphi,\quad
z = \frac{R}{2} (\xi\eta + 1),
\end{equation}
$$
 1 \leq \xi     <    \infty, \qquad
-1 \leq \eta    \leq 1,      \qquad
 0 \leq \varphi <    2\pi,
$$
where $R$ is the interfocus distance.
In the limits where
$R \to 0$ and
$R \to \infty$,
the prolate spheroidal coordinates give back the spherical
coordinates and the parabolic coordinates, respectively
\cite{b20,b30}. In the system of prolate spheroidal coordinates,
the potential $V$ can be written as
\begin{equation}
V = \alpha \frac{2}{R(\xi+\eta)} + \frac{4}{R^2(\xi+\eta)}
\left[ c_1 \frac{1}{(\xi+1)(1+\eta)} +
       c_2 \frac{1}{(\xi-1)(1-\eta)} \right].
\end{equation}
The Schr\"odinger equation (4) for the potential (56)
is separable in prolate
spheroidal coordinates. As a point of fact,
by looking for a solution of
this equation in the form
\begin{equation}
\Psi(\xi, \eta, \varphi) = \psi_{1}(\xi) \psi_{2}(\eta)
{\rm e}^{im\varphi}, \qquad m \in {\bf Z},
\end{equation}
we obtain the two ordinary differential equations
\begin{equation}
\left[ \frac{d}{d\xi}(\xi^2-1) \frac{d}{d\xi} +
\frac{(|m|+\delta_1)^2}{2(\xi+1)} -
\frac{(|m|+\delta_2)^2}{2(\xi-1)} - 2 \alpha R \xi + \frac{ER^2}{2} (\xi^2-1)
\right] \psi_1 = + \lambda(R) \psi_1,
\end{equation}
\begin{equation}
\left[ \frac{d}{d\eta}(1-\eta^2) \frac{d}{d\eta} -
\frac{(|m|+\delta_1)^2}{2(1+\eta)}-
\frac{(|m|+\delta_2)^2}{2(1-\eta)}+ 2 \alpha R\eta + \frac{ER^2}{2} (1-\eta^2)
\right] \psi_2 = - \lambda(R) \psi_2,
\end{equation}
where $\lambda(R)$ is a separation constant in prolate
spheroidal coordinates.
By eliminating the energy $E$ from Eqs. (58) and (59), we
produce the operator
$$
\Lambda  =  \frac{1}{\xi^2-\eta^2} \left[ (1-\eta^2)
\frac{\partial}{\partial\xi} (\xi^2-1) \frac{\partial}{\partial\xi} -
(\xi^2-1)\frac{\partial}{\partial\eta} (1-\eta^2)
\frac{\partial}{\partial\eta} \right] + \frac{2-\xi^2-\eta^2}
{(\xi^2-1)(1-\eta^2)} \frac{\partial^2}{\partial \varphi^2}
$$
\begin{equation}
 - \alpha R \frac{\xi\eta+1}{\xi+\eta} +
2 c_1 \frac{(\xi+\eta)^2 + (\xi-1)(1-\eta)}{(\xi+\eta)(\xi+1)(1+\eta)} +
2 c_2 \frac{(\xi+\eta)^2 - (\xi+1)(1+\eta)}{(\xi+\eta)(\xi-1)(1-\eta)},
\end{equation}
the eigenvalues of which are $\lambda(R)$
and the eigenfunctions of which are $\psi_1(\xi)\psi_2(\eta)$.
The significance of the (self-adjoint) operator
$\Lambda$ can be found by switching to Cartesian coordinates.
A long calculation gives
\begin{equation}
\Lambda =  M + R X,
\end{equation}
where $X$ is the constant of motion (38)
and $M$ is the following constant of motion
\begin{equation}
M = L^2 + c_1 \frac{1}{\cos^2 \frac{\theta}{2}} +
          c_2 \frac{1}{\sin^2 \frac{\theta}{2}},
\end{equation}
the operator $L^{2}$ being the square of the angular momentum operator. The
prolate spheroidal wavefunctions $\psi_1$ and $\psi_2$
could be obtained by solving
(58) and (59). However, it is more economical to proceed
in the following way that shall give us,
 at the same time,
the global wavefunction
(57), i.e., $\Psi(\xi,\eta, \varphi; R; \delta_1, \delta_2)$, and
the interbasis expansion coefficients.

\vspace{0.3cm}
\noindent {\it 6.2. Interbasis Expansions for the Prolate
Spheroidal Wavefunctions}

{}From what preceds, we have three sets of commuting operators,
viz., $\{H, L_z, M\}$, $\{H, L_z, X\}$, and $\{H, L_z,
\Lambda\}$ corresponding to the spherical, parabolic, and
prolate spheroidal coordinates, respectively.
(The operators $L_z$ and $H$ are the $z$-component of the angular momentum
and the Hamiltonian for the generalized Kepler-Coulomb system,
respectively.) In particular, we have
\begin{equation}
M \Psi_{nlm}(r,\theta,\varphi;\delta_1,\delta_2) =
\left( l + \frac{\delta_1+\delta_2}{2} \right)
\left( l + \frac{\delta_1+\delta_2}{2} + 1 \right)
\Psi_{nlm}(r,\theta,\varphi;\delta_1,\delta_2),
\end{equation}
\begin{equation}
X \Psi_{n_1 n_2 m}(\mu,\nu,\varphi;\delta_1,\delta_2) =
- \alpha\frac{ n_1 - n_2 + \frac{\delta_1 - \delta_2}{2} }
{n + \frac{\delta_1+\delta_2}{2}}\,
\Psi_{n_1 n_2 m}(\mu,\nu,\varphi;\delta_1,\delta_2),
\end{equation}
and
\begin{equation}
\Lambda \Psi_{nqm}(\xi,\eta,\varphi; R; \delta_1,\delta_2) =
\lambda_q(R) \Psi_{nqm}(\xi,\eta,\varphi; R; \delta_1,\delta_2)
\end{equation}
for the spherical, parabolic, and prolate spheroidal bases,
respectively. In Eq.~(65), the index $q$ labels the
eigenvalues of the operator $\Lambda$ and varies in the range
$0 \le q \le n - |m| - 1$. We are now ready to deal with the
interbasis expansions
\begin{equation}
\Psi_{nqm}(\xi,\eta,\varphi; R; \delta_1,\delta_2) =
\sum_{l=|m|}^{n-1} T_{nqm}^{l}(R; \delta_1,\delta_2)
\Psi_{nlm}(r,\theta,\varphi;\delta_1,\delta_2),
\end{equation}
\begin{equation}
\Psi_{n   q   m}(\xi,\eta,\varphi; R; \delta_1,\delta_2) =
\sum_{n_1=0}^{n-|m|-1}
                     U_{nqm}^{n_1}(R; \delta_1,\delta_2)
\Psi_{n_1 n_2 m}(\mu, \nu,\varphi;    \delta_1,\delta_2),
\end{equation}
for the prolate spheroidal basis in terms
of the spherical and parabolic bases. [Equation (66)
was first considered by Coulson and Joseph \cite{b31}
in the particular case $\delta_1 =
                        \delta_2 = 0$.]

First, we consider Eq.~(66). Let the operator
$\Lambda$ act on both sides of (66). Then, by using Eqs.~(61),
(63), and (65) as well as the orthonormality property of the
spherical basis, we find that
\begin{equation}
\left[ \lambda_q(R) - \left( l + \frac{\delta_1+\delta_2}{2} \right)
\left( l + \frac{\delta_1+\delta_2}{2} + 1 \right) \right]
T_{nqm}^l(R;\delta_1,\delta_2) =
R \sum_{l'=|m|}^{n-1} T_{nqm}^{l'}(R;\delta_1,\delta_2)
(X)_{ll'},
\end{equation}
where
\begin{equation}
(X)_{ll'} = \int_{0}^{ \infty}
            \int_{0}^{ \pi}
            \int_{0}^{2\pi}
\Psi_{nlm}^*(r,\theta,\varphi;\delta_1,\delta_2)
X \Psi_{nl'm}(r,\theta,\varphi;\delta_1,\delta_2) \,
r^2 \sin\theta dr d\theta d\varphi.
\end{equation}
The calculation of the matrix element $(X)_{ll'}$
can be done by expanding the spherical wavefunctions in (69)
in terms of parabolic wavefunctions [see Eq.~(53)]
and by making use of the eigenvalue
                   equation for $X$ [see Eq.~(64)]. This leads to
\begin{equation}
(X)_{ll'} = - \frac{\alpha}{n + \frac{\delta_1+\delta_2}{2}}
\sum_{n_1=0}^{n-|m|-1}
\left( 2n_1 - n + |m| + \frac{\delta_1-\delta_2}{2} + 1 \right)
\widetilde W_{nlm}^{n_1}(\delta_1,\delta_2)
\widetilde W_{nl'm}^{n_1}(\delta_1,\delta_2).
\end{equation}
Then, by using Eq.~(54) together with the recursion relation \cite{b24}
$$
C_{a\alpha;b\beta}^{c\gamma} =
- \left[ \frac{4c^2(2c+1)(2c-1)}{(c+\gamma)(c-\gamma)(-a+b+c)(a-b+c)
(a+b-c+1)(a+b+c+1)} \right]^{1/2}
$$
$$
\cdot \> \Biggl\{
\left[ \frac{(c-\gamma-1)(c+\gamma-1)(-a+b+c-1)(a-b+c-1)(a+b-c+2)(a+b+c)}
{4(c-1)^2 (2c-3)(2c-1)} \right]^{1/2} C_{a\alpha;b\beta}^{c-2,\gamma}
$$
\begin{equation}
- \frac{(\alpha-\beta)c(c-1) - \gamma a(a+1) + \gamma b(b+1)}
{2c(c-1)} C_{a\alpha;b\beta}^{c-1,\gamma}
\Biggr\}
\end{equation}
and the orthonormality condition
\begin{equation}
\sum_{\alpha,\beta} C_{a\alpha;b\beta}^{c \gamma }
                    C_{a\alpha;b\beta}^{c'\gamma'} =
\delta_{c'      c     }
\delta_{\gamma' \gamma},
\end{equation}
we find that $(X)_{ll'}$
is given by
\begin{equation}
(X)_{ll'} =
\frac{2\alpha}{2n+\delta_1+\delta_2}
\biggl( A_{nm}^{l+1} \delta_{l',l+1} + A_{nm}^l \delta_{l',l-1}
\biggr)-
\frac{\alpha(2|m|+\delta_1+\delta_2)(\delta_1-\delta_2)}
{(2l+\delta_1+\delta_2)(2l+\delta_1+\delta_2+2)} \delta_{l'l},
\end{equation}
where
\begin{equation}
A_{nm}^l = \frac{2}{2l+\delta_1+\delta_2} \left[
\frac{(l-|m|)(l+|m|+\delta_1+\delta_2)(l+\delta_1)(l+\delta_2)(n-l)
(n+l+\delta_1+\delta_2)}{(2l+\delta_1+\delta_2-1)(2l+\delta_1+\delta_2+1)}
\right] ^{1/2}.
\end{equation}
Now by introducing (73) into (68), we get the following
three-term recursion relation for the coefficient $T_{nqm}^l$
$$
\left[
\left(l+\frac{\delta_1+\delta_2}{2}\right)
\left(l+\frac{\delta_1+\delta_2}{2}+1\right)
-\frac{\alpha R(2|m|+\delta_1+\delta_2)(\delta_1-\delta_2)}
{(2l+\delta_1+\delta_2)(2l+\delta_1+\delta_2+2)}
-\lambda_q(R) \right]
                     T_{nqm}^ l   (R ; \delta_1, \delta_2)
$$
\begin{equation}
+ \frac{\alpha R}{2n + \delta_1 + \delta_2}
\left[  A_{nm}^{l+1} T_{nqm}^{l+1}(R ; \delta_1, \delta_2) +
        A_{nm}^ l    T_{nqm}^{l-1}(R ; \delta_1, \delta_2) \right] = 0.
\end{equation}
The recursion relation (75) provides us with a system of $n-|m|$
linear homogeneous equations which can be solved by taking into
account the normalization condition
\begin{equation}
\sum_{l=|m|}^{n-1} | T_{nqm}^l(R; \delta_1,\delta_2) |^2 = 1.
\end{equation}
The eigenvalues $\lambda_{q}(R)$
of the operator $\Lambda$ then follow from the vanishing of the
determinant for the latter system.

 Second, let us concentrate on the expansion (67) of the
prolate spheroidal basis in terms of the parabolic basis.
By employing a technique similar to the one used for deriving
Eq.~(68), we get
\begin{equation}
\left[ \lambda_{q}(R)
 + \alpha R \frac{n_1 - n_2 + \frac{\delta_1-\delta_2}{2}}
{n + \frac{\delta_1+\delta_2}{2}} \right] U_{nqm}^{n_1}(R; \delta_1,\delta_2)
= \sum_{n'_1=0}^{n-|m|-1} U_{nqm}^{n'_1}(R; \delta_1,\delta_2)
(M)_{n_1 n'_1},
\end{equation}
where
\begin{equation}
(M)_{n_1 n'_1} = \int_{0}^{ \infty}
                 \int_{0}^{ \infty}
                 \int_{0}^{2\pi   }
            \Psi_{n _1 n _2 m}^*(\mu,\nu,\varphi;\delta_1,\delta_2)
M           \Psi_{n'_1 n'_2 m}  (\mu,\nu,\varphi;\delta_1,\delta_2)
  \, \frac{\mu + \nu}{4} d\mu d\nu d\varphi.
\end{equation}
The matrix elements $(M)_{n_1 n_1'}$ can be calculated
in the same way as  $(X)_{l   l'  }$ except that now we
must use the relation \cite{b24}
$$
[(b-a+c)(a-b+c+1)]^{1/2} C_{a\alpha;b\beta}^{c\gamma} =
[(a-\alpha+1)(b-\beta)]^{1/2}
C_{a+1/2, \alpha-1/2; b-1/2, \beta+1/2}^{c\gamma}
$$
\begin{equation}
+ [(a+\alpha+1)(b+\beta)]^{1/2}
C_{a+1/2, \alpha+1/2; b-1/2, \beta-1/2}^{c\gamma}
\end{equation}
and the orthonormality condition
\begin{equation}
\sum_{c, \gamma} C_{a\alpha ;b\beta }^{c\gamma}
                 C_{a\alpha';b\beta'}^{c\gamma}
= \delta_{\alpha\alpha'} \delta_{\beta\beta'},
\end{equation}
instead of Eqs.~(71) and
                (72). This yields the matrix element
$$
(M)_{n_1 n'_1}=\biggl[(n-n_1-|m|-1)(n_1+1)+(n_1+|m|+\delta_1)
(n-n_1+\delta_2)
+\frac{1}{4}(\delta_1-\delta_2)(\delta_1-\delta_2-2)
\biggr] \delta_{n'_1 n_1}
$$
$$
- [(n-n_1-|m|-1)(n_1+1)(n_1+|m|+\delta_1+1)(n-n_1 + \delta_2-1)]^{1/2}
\,\delta_{n'_1, n_1+1}
$$
\begin{equation}
- [ n_1(n-n_1-|m|)(n_1+|m|+\delta_1)(n-n_1+\delta_2)]^{1/2}
\,\delta_{n'_1, n_1-1}.
\end{equation}
Finally, the introduction of (81) into (77)
leads to the three-term recursion relation
$$
\Biggl[ (n-n_1-|m|-1)(n_1+1)(n_1+|m|+\delta_1)(n-n_1+\delta_2) +
\frac{1}{4} (\delta_1-\delta_2)(\delta_1-\delta_2-2)
$$
$$
- \alpha R \,
\frac{2 n_1 - n + |m| + 1 + \frac{\delta_1-\delta_2}{2}}
{n + \frac{\delta_1+\delta_2}{2}} - \lambda_{q}(R) \Biggr]\,
U_{nqm}^{n_1}(R; \delta_1,\delta_2) =
$$
$$
[(n-n_1-|m|-1)(n_1+1)(n_1+|m|+\delta_1+1)(n-n_1+\delta_2-1)]^{1/2}
\,U_{nqm}^{n_1+1}(R; \delta_1,\delta_2)
$$
\begin{equation}
+ [ n_1(n-n_1-|m|)(n_1+|m|+\delta_1)(n-n_1+\delta_2)]^{1/2}
U_{nqm}^{n_1-1}(R; \delta_1,\delta_2)
\end{equation}
for the expansion coefficients $U_{nqm}^{n_1}(R;
\delta_1,\delta_2)$. This relation can be iterated by taking
account of the normalization condition
\begin{equation}
\sum_{n_1=0}^{n-|m|-1}| U_{nqm}^{n_1}(R; \delta_1,\delta_2) |^2 = 1.
\end{equation}
Here again, the eigenvalues $\lambda_q(R)$ may be obtained by
solving a system of $n - |m|$ linear homogeneous equations.

In the case $\delta_1 = \delta_2 = 0$, from Eqs.~(75) and (82)
we obtain
three-term recursion relations for the coefficients of
interbasis
expansions of the prolate spheroidal basis in spherical and parabolic
bases for the hydrogen atom~; these coefficients were
calculated in Refs.~[28,~44,~45].

Finally, it should be noted
that the following four limits
\begin{equation}
\lim_{R \rightarrow \infty}
 U_{nqm}^{n_1}(R; \delta_1,\delta_2) = \delta_{n_1 q}, \qquad
\lim_{R \rightarrow 0}
 U_{nqm}^{n_1}(R; \delta_1,\delta_2) = {\widetilde W}_{n q m}^{n_1}
(\delta_1, \delta_2),
\end{equation}
\begin{equation}
\lim_{R \rightarrow 0}
 T_{nqm}^{l}  (R; \delta_1,\delta_2) = \delta_{l   q}, \qquad
\lim_{R \rightarrow \infty}
 T_{nqm}^{l}  (R; \delta_1,\delta_2) =             W _{n q m}^l
(\delta_1, \delta_2)
\end{equation}
furnish a useful means for checking
the calculations presented in Sections 5 and 6.

\begin{center}
\section{.~Concluding Remarks}
\end{center}

One of the main results of this paper concerns the construction
of the prolate spheroidal
basis for the generalized Kepler-Coulomb system (corresponding
to $\delta_1 \ne 0$ and
   $\delta_2 \ne 0$) as
a superposition of either the spherical basis or the parabolic
basis. In this respect, our work is a continuation of the
pioneer work by Coulson and Joseph \cite{b31}
on the ordinary
Kepler-Coulomb system (corresponding
to $\delta_1 = 0$ and
   $\delta_2 = 0$). We have obtained three-term recursion
relations for the coefficients of the interbasis expansions
of the prolate spheroidal basis in terms of the spherical basis,
on one hand, and of the parabolic basis, on the other hand.
These recursion relations can be easily implemented by using
symbolic programming languages (like MAPLE, MATHEMATICA, etc.)
especially, in the framework of perturbation theory, for small
or large values of the spheroidal separation constant $R$.

We have also calculated the coefficients of interbasis
expansions between the spherical basis and the parabolic basis.
As we could expect from the limiting case $\delta_1 =
                                           \delta_2 =0$, the
latter coefficients are Clebsch-Gordan
coefficients for the special unitary group SU(2) modulo an analytic
continuation
to real values of their arguments.

It was realized long time ago that prolate spheroidal
coordinates may be useful in quantum chemistry even
in situations exhibiting spherical
symmetry [28,~29]. This work shows the relevance of
prolate spheroidal
coordinates in axial symmetry. This is indeed the kind of
symmetry that is encountered in the theory of the chemical
bond.

We close this paper with a remark of interest for
mathematical physics (and
mathematical chemistry). Most of the $\delta_1$-
                                 and $\delta_2$-dependent expressions
(e.g., the ring-shaped functions)
in this
work present nice properties with respect to the arguments
 $\delta_1$ and
 $\delta_2$. It would be worth to study some of these
expressions in connection with the theory of orthogonal
polynomials.

\vspace{0.3cm}
\begin{center}
{\Large\bf Acknowledgments}
\end{center}

The present collaboration profited from a visit of G.S.P. to
the {\it Universit\'e Claude Bernard Lyon-1}. He is grateful
to the {\it Institut de Physique Nucl\'eaire de Lyon}
for the hospitality extended to him during his stay
in Lyon-Villeurbanne.

\vfill\eject

\begin{center}
{\Large\bf Appendix: An Alternative Proof for the Bi-Orthogonality
of the Radial Wavefunctions}
\end{center}

We start from two copies of Eq.~(15): one for the pair
($E\equiv E_{n}, R\equiv R_{nl}$)
and the other one for ($E\equiv E_{n'}, R\equiv R_{n'l'}$).
After multiplication of the first copy (respectively second
copy) by $R_{n'l'}$ (respectively $R_{nl}$) and integration,
with the measure $r^2 dr$, on the half real line, we obtain
$$
(l'-l)(l+l'+\delta_1+\delta_2+1)
\int_{0}^{\infty} R_{n'l'}(r;\delta_1,\delta_2) R_{nl}(r;\delta_1,\delta_2)
\, dr
$$
\begin{equation}
 =  2(E_n - E_{n'})
\int_{0}^{\infty} R_{n'l'}(r;\delta_1,\delta_2) R_{nl}(r;\delta_1,\delta_2)
r^2 \, dr.
\end{equation}
{}From the latter equation, it readily follows that
\begin{equation}
\int_{0}^{\infty} R_{n'l}(r;\delta_1,\delta_2) R_{nl}(r;\delta_1,\delta_2)
r^2 dr  = 0 \,\,\,\,\,\,\, {\rm for}\,\, n'\neq n
\end{equation}
and
\begin{equation}
\int_{0}^{\infty} R_{nl'}(r;\delta_1,\delta_2)
                  R_{nl }(r;\delta_1,\delta_2)
\,\, dr = 0 \,\,\,\,\,\,\, {\rm for} \,\, l' \neq l.
\end{equation}
This completes the proof of the two orthogonality
relations $I_{nn'}\sim \delta_{nn'}$ and $J_{ll'}\sim
\delta_{ll'}$.
It should be emphasized that Eq.~(87) turns out
to be a direct consequence of the independence
of the energy $E$ on the quantum number $l$.

\vfill\eject

\end{document}